\documentclass[oldversion]{aa}
\usepackage{graphicx}
\usepackage{eufrak}
\usepackage{txfonts}
\usepackage{natbib}
\bibpunct{(}{)}{;}{a}{}{,}

\begin{document}

\title{The origin of carbon:\\ Low-mass stars and an evolving, initially top-heavy IMF?}
\titlerunning{The origin of carbon}

\author{Lars Mattsson\thanks{\email{mattsson@fysast.uu.se}}}
\institute{Dept. Physics and Astronomy, Div. of Astronomy and Space Physics, Uppsala University, Box 515, SE-751 20 Uppsala, Sweden}

\offprints{Lars Mattsson}

\date{Received date; accepted date}

\abstract{Multi-zone chemical evolution models (CEMs), differing in the nucleosynthesis prescriptions (yields) and prescriptions of star formation, 
have been computed for the Milky Way. All models fit the observed O/H and Fe/H gradients well and reproduce the main characteristics of the gas 
distribution, but they are also designed to do so. For the C/H gradient the results are inconclusive with regards to yields and star formation.
The C/Fe and O/Fe vs. Fe/H, as well as C/O vs. O/H trends predicted by the models for the solar neighbourhood zone were compared with stellar 
abundances from the literature. For O/Fe vs. Fe/H all models fit the data, but for C/O vs. O/H, only models with increased carbon yields for
zero-metallicity stars or an evolving initial mass function provide good fits. Furthermore, a steep star formation threshold in the disc can be ruled
out since it predicts a steep fall-off in all abundance gradients beyond a certain galactocentric distance ($\sim 13$ kpc) and cannot explain the
possible flattening of the C/H and Fe/H gradients in the outer disc seen in observations. Since in the best-fit models the enrichment scenario is such 
that carbon is primarily produced in low-mass stars, it is suggested that in every environment where the peak of star formation happened a few
Gyr back in time, \it winds of carbon-stars \rm are responsible for most of the carbon enrichment. However, a significant contribution by
zero-metallicity stars, especially at very early stages, and by winds of high-mass stars, which are increasing in strength with metallicity,
cannot be ruled out by the CEMs presented here. In the solar neighbourhood, as much as 80\%, or as little as
40\% of the carbon may have been injected to the interstellar medium by low- and intermediate-mass stars. 
The stellar origin of carbon remains an open question, although production in low- and intermediate-mass stars appears to be the simplest 
explanation of observed carbon abundance trends.}


\maketitle

\section{Introduction}
Carbon is one of the most common elements,
but we know surprisingly little about its origin. What we do know, however, is that it is ubiquitous throughout the Universe, and
can be found in just about any astrophysical environment. We also know that life, as we know it, requires the existence of carbon, nitrogen,
oxygen and a few other elements. Understanding the origin of carbon may therefore tell us something about the probability of finding carbon-based
life elsewhere in the Galaxy, i.e., beyond the solar neighbourhood.

The stellar origin of carbon is mainly due to the Triple-Alpha reaction \cite{Salpeter52} but this reaction may occur in various types of stars.
Carbon Stars (C-stars) have been recognised as a class of astronomical object for more than a century and have several times been suggested as the
main carbon sources in the Universe. Already in the work by Burbidge et al. (1957) \nocite{Burbidge57} it was suggested that carbon was provided by 
mass-loss from red giants and supergiants. Later Dearborn (1978) \nocite{Dearborn78} suggested that low-mass stars may be a significant source of 
carbon {based on abundance determinations} in planetary nebulae. Recent theoretical work on stellar evolution of low and intermediate mass (LIM) 
stars also {points to} low-mass stars as {being} significant producers of carbon, although the quantative results may differ 
\cite{Marigo01,Izzard04,Gavilan05,Karakas07}.

Models of chemical evolution (CEMs) are in general in good agreement with observed abundances if a delayed carbon release from LIM-stars 
is assumed. Timmes et al. (1995) \nocite{Timmes95} used the nucleosynthetic yields by Woosley \& Weaver (1995, from hereon cited as WW95) and 
Renzini \& Voli (1981), which led to a very significant contribution of carbon from LIM-stars to the Galactic Disc, and more recent work (using other 
sets of yields for LIM-stars) have led to quite similar results \cite{Chiappini03,Akerman04,Carigi05,Gavilan05}. However, Maeder (1992) 
\nocite{Maeder92} argued that radiatively driven winds from high-mass (HM) stars should provide huge amounts of helium and carbon. In such
case, these stars would be the main contributors, and in CEMs the {role} of LIM-stars would have to be much less significant 
to avoid over-production of carbon compared with observed abundances. Garnett et al. (1995) \nocite{Garnett95} observed that the C/O-ratio
increased with increasing O/H in dwarf irregular galaxies, which they interpreted as {being} consistent with carbon being produced in {HM-stars} 
with metallicity-dependent yields, as in the models by Maeder (1992) and Portinari et al. (1998). Following that idea, {Gustafsson et al.~(1999)} 
argued that the rising C/O-trend with metallicity that they found in Galactic-disc stars was the result of carbon being produced in HM-stars rather 
than LIM-stars.

Recent observations have revealed a declining trend {with increasing metallicity} for the C/O ratio in the solar neighbourhood at early times 
\cite{Akerman04,Fabbian09}. C/O vs. O/H shows a negative slope roughly until the anticipated onset of disc formation, i.e., during the first billion 
years of Galactic evolution, which is in disagreement \cite[see, e.g.][]{Chiappini03,Gavilan05} with the predictions of CEMs {that} do not include any 
modifications of the {standard} carbon and oxygen yields. {Such C/O discrepancy is obviously} connected to the first generations of stars in the Milky 
Way, and may therefore have {common origin with our still incomplete understanding of} early chemical evolution. For instance, the
underabundance of essentially metal-free LIM-stars in the halo, which is often claimed to be a result of a top-heavy initial mass function (IMF) at
early times \cite[see, e.g.,][and references therein]{Abel02,Tumlinson06,Karlsson08}, is one such problem. If the IMF has evolved from being 
initially top-heavy, to the form that is observed in the solar neighbourhood today, it is possible that this may affect the evolution of the C/O 
ratio as well. Chiappini et al. (2000) have considered an evolving IMF, but concluded that it did not improve the agreement with observational 
constraints. However, their study did not focus on the early chemical evolution and abundance trends at very low metallicity.

In this paper, the origin of carbon is investigated once again, using a set of multi-zone CEMs for the Milky Way. Several nucleosynthetic
prescriptions are considered, as well as the effects of an evolving IMF.

\begin{figure}
\resizebox{\hsize}{!}{
\includegraphics{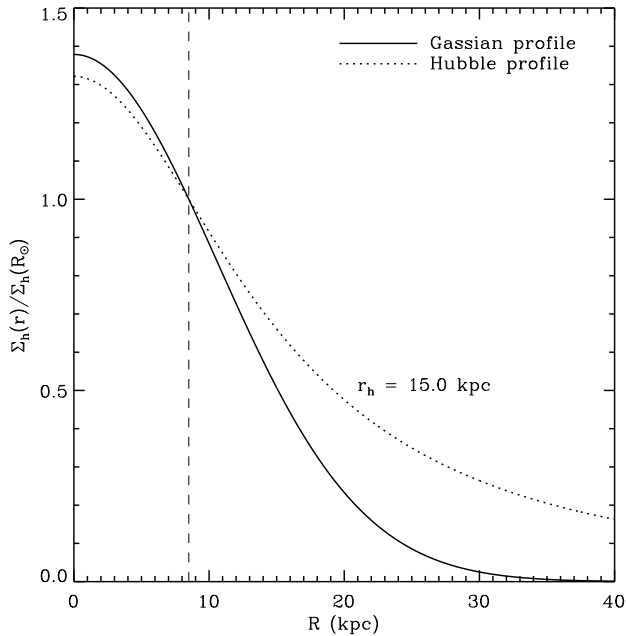}}
\caption{\label{halo}
The gassian surface density profile adopted for the halo/thick disc in this paper compared to a Hubble (1923) profile, normalised to the
{density at the Sun's galactocentric distance (marked by a vertical dashed line in the figure)}.}
\end{figure}

\section{Chemical evolution model}
\label{cem}

\subsection{Galaxy Formation}
It is usually assumed that the Galaxy was formed through baryonic infall, or more precisely, by accretion of pristine gas (hydrogen and helium),
and that the rate of accretion follows an exponential decay \cite{Lacey85,Timmes95}.
Furthermore, it is rather well established that the halo/thick disc and the thin disc components of the Galaxy were assembled on different time-scales
and, perhaps, with some separation in time, as suggested by Chiappini et al. (1997). The latter scenario, known as the {\it two-infall model},
consists of two infall episodes where the disc formation starts after some time $\tau_{\rm max}$ (which is {when} the rate of accretion
reaches its maximum), i.e., the total rate of accretion is
\begin{equation}
\dot{\Sigma}_{\rm inf.}(r,t) = \left\{
\begin{array}{lll}
\dot{\Sigma}_{\rm h}(r,t) & \mbox{if} & t \le \tau_{\rm max}\\
\dot{\Sigma}_{\rm h}(r,t) + \dot{\Sigma}_{\rm d}(r,t) & \mbox{if} & t > \tau_{\rm max}
\end{array}
\right.
\end{equation}
where, for the halo/thick disc,
\begin{equation}
\label{infall}
\dot{\Sigma}_{\rm h}(r,t) = {\Sigma_{\rm h}(r,t_0)\over\tau_{\rm h}}
\left\{1-\exp\left[-{t_0\over \tau_{\rm h}}\right]\right\}^{-1}\exp\left[-{t\over\tau_{\rm h}}\right],
\end{equation}
and for the (thin) disc,
\begin{eqnarray}
\label{infall2}
\nonumber
\dot{\Sigma}_{\rm d}(r,t) =
{\Sigma_{\rm d}(r,t_0)\over\tau_{\rm d}(r)}\left\{1-\exp\left[-{(t_0-\tau_{\rm max})\over \tau_{\rm d}(r)}\right]\right\}^{-1}&\\
\times \exp\left[-{(t-\tau_{\rm max})\over\tau_{\rm d}(r)}\right]&,
\end{eqnarray}
where $\tau$ is the infall time scale and $t=t_0$ is the age of the Galaxy. Here, the value used for $t_0$ is 13.5 Gyr, which can be regarded
as an upper limit. To simulate the inside-out formation of a galactic disc, the infall time scale is assumed to have a linear radial dependence
\cite[for further details, see][and references therein]{Chiosi80,Matteucci89},
\begin{equation}
\tau_{\rm d}(r) = {\rm max}\left[0, \tau_\odot - \tau_0\left(1-{r\over R_\odot}\right)\right],
\end{equation}
where $\tau_\odot$ is the infall time scale in the solar neighbourhood, $\tau_0$ is an arbitrary {constant} and $R_\odot = 8.5$ kpc is the
galactocentric distance {of} the Sun. For the halo/thick-disc phase it is assumed that $\tau_{\rm h} = 1.0$ Gyr at all galactocentric distances.

The {present-day total baryon density in the solar neighbourhood is assumed to be $60M_\odot$~pc$^2$
\cite[which is consistent with the results obtained by, e.g.,][]{Holmberg04} and the} final baryonic
(thin) disc is assumed to follow an exponential distribution,
\begin{equation}
\Sigma_{\rm d}(r,t_0) = \Sigma_0(t_0) \exp\left(-{r\over r_{\rm d}}\right),
\end{equation}
where $\Sigma_0$ is the present-day central surface density and $r_{\rm d}$ is the disc scale length. The value of {the latter} parameter is not
very well constrained for the Milky Way. In this study it is assumed that $r_{\rm d}$ is rather short, which is supported by stellar statistics
\cite{Ruphy96,Porcel98}.
The final density $\Sigma_{\rm h}$ of baryonic matter of the halo/thick-disc component in the solar neighbourhood
is of the order of $10 M_\odot$ pc$^{-2}$. {The present-day density of halo stars in the solar neighbourhood has been estimated
to $5.7 \cdot 10^{-4} M_\odot$ kpc$^{-3}$ \cite{Preston91}, which corresponds to a surface density of $1-2 M_\odot$ pc$^{-2}$ and for the thick
disc stars the local surface density is about 3.5\% of that of the thin disc \cite{Ohja01}. But $\Sigma_{\rm h}$ remains difficult to constrain since the
corresponding gas fraction is currently unknown.} The halo/thick-disc surface density is modelled by a gaussian,
\begin{equation}
\Sigma_{\rm h} (r,t) = \Sigma_{\rm h}(R_\odot,t_0)\,\exp\left(-{r^2-R_\odot^2\over r_{\rm h}^2}\right),
\end{equation}
where $R_\odot$ is defined as above and $r_{\rm h}$ is the scale length. The inner part of this profile is similar to a Hubble (1923) profile, used
in some other studies \cite[e.g.,][]{Renda05}, while the outer part is gradually steeper
(see Fig. \ref{halo}), as suggested by very deep star counts in the halo \cite[see][and references therein]{Helmi08}. For further details, see Table
\ref{parameters}.

  \begin{table*}
  \begin{center}
  \caption{\label{parameters} Parameters for the CEMs.}
  \begin{tabular}{lllllllllll}
  \hline
  \hline
  Model     & $\Sigma_{\rm h}(R_\odot,t_0)$ & $r_{\rm h}$ & $\Sigma_{\rm d}(R_\odot,t_0)$ & $r_{\rm d}$  & $\nu_{\rm h}$ & $\nu_{\rm d}$ & $\nu'_{\rm d}$ & $\varepsilon$ & $m_{\rm u}$ & $m_{\rm c}$ \\
            & $M_\odot$ pc$^{-2}$           & kpc         & $M_\odot$ pc$^{-2}$           & kpc          & Gyr$^{-1}$    & Gyr$^{-1}$    & Gyr$^{-1}$     &               & $M_\odot$   & $M_\odot$   \\
  \hline
  A1        & 10.0                          & 15.0        & 50.0                          & 2.1          & 0.170         & 0.170         & 0.0425         & 1.5           & 112.0       & 0.32        \\
  B1        & 10.0                          & 15.0        & 50.0                          & 2.1          & 0.138         & 0.138         & 0.0345         & 1.5           & 100.0       & 0.40        \\
  C1        & 10.0                          & 15.0        & 50.0                          & 2.1          & 0.163         & 0.163         & 0.0408         & 1.5           & 113.0       & 0.32        \\
  D1        & 10.0                          & 15.0        & 50.0                          & 2.1          & 0.166         & 0.166         & 0.0415         & 1.5           & 109.0       & 0.32        \\
  E1        & 10.0                          & 15.0        & 50.0                          & 2.1          & 0.176         & 0.176         & 0.0440         & 1.5           & -           & -           \\[2mm]
  A2        & 15.0                          & 11.5        & 45.0                          & 2.1          & 0.535         & 0.535         & 0.0000         & 1.0           & 114.0       & 0.30        \\
  B2        & 15.0                          & 11.5        & 45.0                          & 2.1          & 0.404         & 0.404         & 0.0000         & 1.0           & 100.0       & 0.40        \\
  C2        & 15.0                          & 11.5        & 45.0                          & 2.1          & 0.535         & 0.535         & 0.0000         & 1.0           & 114.0       & 0.30        \\
  D2        & 15.0                          & 11.5        & 45.0                          & 2.1          & 0.535         & 0.535         & 0.0000         & 1.0           & 110.0       & 0.30        \\
  \hline
  \end{tabular}
  \end{center}
  \end{table*}

\subsection{Star formation}
For the halo/thick disc, the star-formation rate is prescribed by a modified Schmidt-law of the form
\begin{equation}
\label{sfr}
\dot{\Sigma}_\star(r,t) = \nu_{\rm h}\,\Sigma(r,t)\,
\,\left[{\Sigma_{\rm gas}(r,t)\over \Sigma(r,t)} \right]^{1+\varepsilon},
\end{equation}
where $\nu_{\rm h}$ is the star-formation efficiency expressed in Gyr$^{-1}$. For the halo/thick disc phase it is assumed that $\nu_{\rm h}$ is
constant, while for the disc phase the star-formation efficiency is assumed to be related to the angular frequency of the disc
\cite{Wyse89,Boissier03}, i.e.,
\begin{equation}
\label{sfre}
\dot{\Sigma}_\star(r,t) =
\nu (r,t)\,\Sigma(R_\odot,t_0)\,\left[{\Sigma_{\rm gas}(r,t)\over \Sigma(R_\odot,t_0)} \right]^{1+\varepsilon},
\end{equation}
and
\begin{equation}
\nu(r,t) = \left\{
\begin{array}{lll}
\nu_{\rm d}\,\biggl{[\displaystyle{\Omega_0(r)\over \Omega_{0}(R_\odot)}\biggr]} & \mbox{if} & \Sigma_{\rm gas} \ge \Sigma_{\rm c},\\[5mm]
\nu'_{\rm d} & \mbox{if} & \Sigma_{\rm gas}  <  \Sigma_{\rm c},
\end{array}
\right.
\end{equation}
where {$\Sigma_{\rm gas}$ is the gas density, $\Sigma_{\rm c}$ is the critical density,}
$\nu_{\rm d}$, $\nu'_{\rm d}$ are constants and $\Omega_0$ is the present-day ($t=t_0$) angular frequency of the disc (rotation {data} were
taken from Sofue et al. 1999). To avoid unphysical "kinks" when switching between the two star formation regimes, a "smooth step
function", defined as,
\begin{equation}
\theta(r,t) \equiv {1\over 1+ \exp\left[-2k\,(\Sigma_{\rm gas} - \Sigma_{\rm c})\right]},
\end{equation}
is used instead of an actual conditional expression. This function will approach a regular Heaviside step function as $k\to \infty$, but here it is
assumed that $k=1$. The critical density $\Sigma_{\rm c}$ may change along the disc, but for simplicity it is assumed that
$\Sigma_{\rm c} = 7.0 M_\odot$~pc$^{-2}$ over the whole disc \cite{Kennicutt89}. Two different
star formation prescriptions are considered in this paper: (1) a Schmidt law as above with $\varepsilon = 0.5$ \cite[cf.][]{Fuchs09} above the
critical gas density $\Sigma_{\rm c}$, and {below $\Sigma_{\rm c}$ star formation proceeds} with lower efficiency according to a linear Schmidt-law
$\dot{\Sigma}_\star = \nu'_{\rm d}\,\Sigma_{\rm gas}$, where $\nu'_{\rm d} = 0.25 \nu_{\rm d}$ {(models A1 to E1 in Table \ref{parameters})},
and (2) a Schmidt law with $\varepsilon = 0$ and a strict threshold, i.e., $\nu'_{\rm d} = 0 \to \dot{\Sigma}_\star = 0$ below the critical density
{(models A2 to D2 in Table \ref{parameters})}. The first case will be referred to as star formation law of type 1, and the second case as type 2.

\subsection{The stellar initial mass function}
\begin{figure}
\resizebox{\hsize}{!}{
\includegraphics{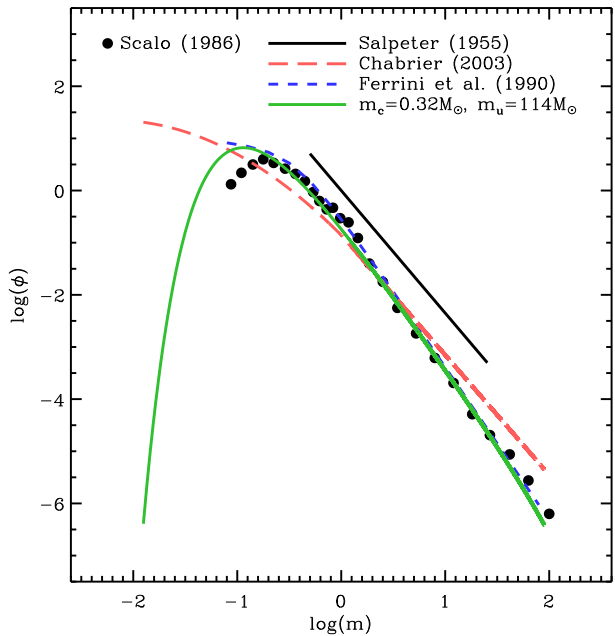}}
\caption{\label{imf}
Comparison between the IMF $\phi(m)$ used in this paper and some IMFs frequently used in the literature. {The stellar mass $m$ is
given in solar masses.}}
\end{figure}

It is assumed that stars are formed according to a stellar IMF that may, or may not, be time-dependent. In its simplest
form, the IMF is just a power-law with sharp cut-offs at the lower and upper ends. It is well established, however, that the IMF turns over at low
masses and is probably truncated at the high-mass end (see Fig. \ref{imf} and references therein). Hence, an IMF of the form
\begin{equation}
\label{imfunc}
\phi(m,t)=\phi_0(t)\,m^{-(1+x)}\exp\left\{-\left[{m_{\rm c}(t)\over m} + {m\over m_{\rm u}(t)} \right]\right\},
\end{equation}
is adopted, where $m_{\rm c}$, $m_{\rm u}$ are the masses defining the low-mass turn over and the high-mass truncation of the IMF, respectively.
The two parameters $m_{\rm c}$, $m_{\rm u}$ may be regarded as functions of time, thus allowing for an evolving IMF. The constant $\phi_0$, which
is obtained by the normalisation condition
\begin{equation}
\int_{0}^{\infty} m\,\phi(m,t)\,dm = 1,
\end{equation}
is then also a function of time. More precisely,
\begin{equation}
\phi_0(t) = {1\over 2}\,{\mu^{x}(t)\over K_{x-1}(\mu)}, \quad x \ge 1,
\end{equation}
where $\mu$ is a dimensionless variable defined as
\begin{equation}
\mu(t) \equiv \sqrt{{m_{\rm c}(t)\over m_{\rm u}(t)}},
\end{equation}
and $K_n$ is the modified Bessel function of the second kind and order $n$. The mean stellar mass of this IMF is
\begin{equation}
\langle m\rangle (t) = m_{\rm u}(t)\,\mu(t) \,{K_{x-1}(\mu) \over K_{x}(\mu)}, \quad x \ge 1,
\end{equation}
while the most probable mass is given by
\begin{equation}
m_0(t) = {1\over 2} x\,m_{\rm u}(t)\left[\sqrt{1+ 4x^2\mu(t)^2} - 1\right].
\end{equation}
Both the mean and most probable masses are also functions of time in the general case. However, in the special case when
$m_{\rm c}(t) \propto m_{\rm u}(t)$, or $\mu = \mbox{const.}$, the situation is somewhat simpler. First of all, $\phi_0$ becomes constant
over time and the IMF {does not need} to be renormalised at every time step. This simplifying assumption will be used throughout the following.

\begin{figure}
\resizebox{\hsize}{!}{
\includegraphics{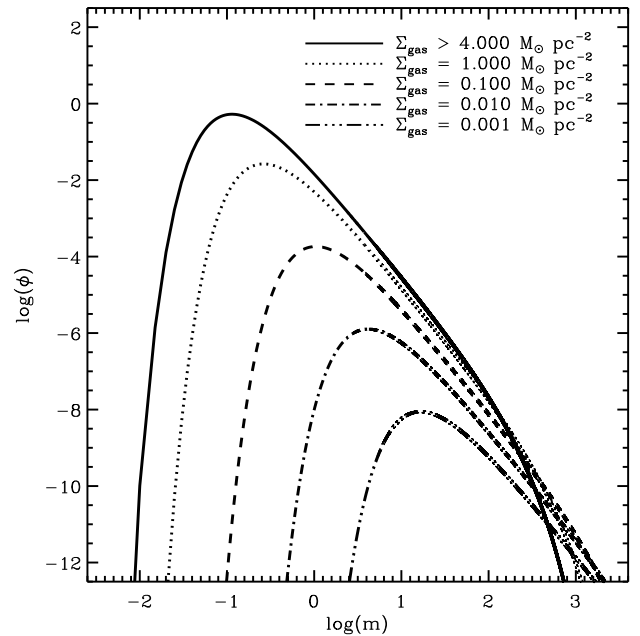}}
\caption{\label{imfevol}
Evolution of the IMF with the gas density in model E1. {Meaning of the axes as in Fig. \ref{imf}.}}
\end{figure}

The time-evolution of the parameters $m_{\rm c}$ and $m_{\rm u}$ cannot be completely arbitrary. They are quite likely related to some
characteristic mass-scale related to the physical origin of the IMF. Larson (1995; 1996; 1998) pointed out that there should be a connection
between the turn-over mass (or the characteristic stellar mass) and some fundamental mass-scale in the star formation process, such as the Jeans
mass \cite{Jeans02}. Recent observational evidence for an evolving IMF {suggests} characteristic stellar masses which
are consistent with such a picture \cite{vanDokkum08}. The thermal Jeans mass is given by the pressure and temperature of the
collapsing cloud, i.e.,  $m_{\rm J} \propto T^2 P^{-1/2}$. In a self-gravitating cloud, the pressure-gravity balance is such that the Jeans mass
can also be expressed in terms of the gas surface density $\Sigma$ as $m_{\rm J} \propto T^2 \Sigma^{-1}$ (see, e.g., Larson 1985). If the ISM is
isothermal, the Jeans mass is inversely proportional to the local surface density of the gas and if cooling is inefficient, it may be
close to constant {in time}. In the following {we assume} that the gas temperature $T$ is a function of the gas density $\Sigma$ alone, i.e., a
polytropic equation of state. Assuming then that $m_{\rm c}$ and $m_{\rm u}$ are related to the characteristic Jeans mass of the ISM,
it is reasonable to parametrise these quantities in {terms} of the local gas mass density of the Galactic disc \cite{Elmegreen99}. In
particular, a power-law form, $m_{\rm c}(t) \propto m_{\rm u}(t) \propto \Sigma_{\rm gas}^{-\beta}(t)$, with $\beta \in [0,1]$, {gives an
adequate description of} how $m_{\rm c}$ and $m_{\rm u}$ change during the evolution of the Galaxy {(see Fig. \ref{imfevol}, showing
how the IMF changes with the gas density)}.

In the present study, two types of IMFs according to Eq. (\ref{imfunc}) are considered:
\begin{itemize}
\item[(a)] A non-evolving IMF {of the form}
\begin{equation}
\phi(m)=\phi_0\,m^{-(1+x)}\exp\left[-\left({m_{\rm c}\over m} + {m\over m_{\rm u}} \right)\right],
\end{equation}
where the upper mass-cut is $100-120M_\odot$ in order to avoid over-production of oxygen, and the turn-over
mass $m_{\rm c}$ is $0.30-0.40 M_\odot$. The power-law index $x$ was chosen to be steeper than
the canonical \cite{Salpeter55} value, i.e., $x = 1.80$ instead of $x = 1.35$, in order to reproduce the properties and abundances of the solar
neighbourhood (see, e.g., Chiappini et al. 1997).
\item[(b)] A time-dependent IMF {of the form given in Eq. (\ref{imfunc})}, where
\begin{equation}
\label{evolimf}
m_{\rm c}(t) = \left\{
\begin{array}{lll}
m_{\rm c}(t_0)\,\biggl(\displaystyle{\Sigma_{\rm gas}\over 4M_\odot\mbox{pc}^{-2}}\biggl)^{-0.6}\, & \mbox{if} & \Sigma_{\rm gas} \le 4M_\odot\mbox{pc}^{-2}\\[5mm]
m_{\rm c}(t_0)\, & \mbox{if} & \Sigma_{\rm gas} > 4M_\odot\mbox{pc}^{-2}
\end{array}
\right. ,
\end{equation}
and
\begin{equation}
m_{\rm u}(t) = \left\{
\begin{array}{lll}
m_{\rm u}(t_0)\,\biggl(\displaystyle{\Sigma_{\rm gas}\over 4M_\odot\mbox{pc}^{-2}}\biggl)^{-0.6}\, & \mbox{if} & \Sigma_{\rm gas} \le 4M_\odot\mbox{pc}^{-2}\\[5mm]
m_{\rm u}(t_0)\, & \mbox{if} & \Sigma_{\rm gas} > 4M_\odot\mbox{pc}^{-2}.
\end{array}
\right.
\end{equation}
In the equations above, $m_{\rm c}(t_0)$ and $m_{\rm u}(t_0)$ denote the present-day values for the local-disc IMF.
Note that when $\Sigma_{\rm gas} > 4 M_\odot$ pc$^{-2}$, this IMF is nearly identical to that of case (a).
\end{itemize}

A comparison between the IMF adopted here and a few IMFs commonly used in CEMs is shown in Fig. \ref{imf}. Note that {case (b)} above will lead
to a larger fraction of very low-mass stars and substellar objects, compared to, e.g., the {IMF by Scalo (1986)}, which means that a larger
fraction of metals will be locked-up in stellar objects that do not release any significant amounts of gas back to the ISM within a Hubble time.

In order to include the contribution from supernovae type Ia (SNIa) events, which is commonly assumed to be the result of gas accretion onto a white
dwarf from its companion (presumably a red giant) in a binary system, the formalism introduced by Greggio \& Renzini (1983) \nocite{Greggio83} and
Matteucci \& Greggio (1986) \nocite{Matteucci86} is used. In this formalism,
the equation describing the evolution of an element $i$ can be written as
\begin{eqnarray}
\nonumber
{\partial\Sigma_i\over \partial t} &=& \,\dot{\Sigma}_{{\rm inf},i}(r,t) - \frac{\Sigma_i(r,t)}{\Sigma(r,t)}\,\dot{\Sigma}_{\star}(r,t)+\\
\nonumber
&&\,\int_{16}^{\infty} X_i(r, t-\tau_m)\,\phi(m)\,\dot{\Sigma}_{\star}(r,t-\tau_m)\,dm+\\
&&\,\int_{0}^{3} X_i(r, t-\tau_m)\,\phi(m)\,\dot{\Sigma}_{\star}(r,t-\tau_m)\,dm+\\
\nonumber
&&\,(1-\eta)\int_{3}^{16} X_i(r, t-\tau_m)\,\phi(m)\,\dot{\Sigma}_{\star}(r,t-\tau_m)\,dm+\\
\nonumber
&&\,\eta\int_{3}^{16} \int_{\mu_{\rm min}}^{1/2} \varphi(\mu)
X_i(r, t-\tau_m)\,\phi(m)\,\dot{\Sigma}_{\star}(r,t-\tau_m)\,d\mu\,dm,
\end{eqnarray}
where $\tau_m$ is the {lifetime} of a star of mass $m$ (as obtained from the stellar evolutionary tracks by Schaller et al. 1992 and WW95),
$\phi(m)$ is the IMF described above, $X_i$ is the production matrix for an element $i$ as defined by Talbot \& Arnett (1971, 1973) and $\eta$ is
the fraction of (binary) low and intermediate mass stars undergoing a SNIa event. In the equation above, it is implicitly assumed that SNIa events
originate from binary systems more massive than $3 M_\odot$ and with a maximum mass of $16 M_\odot$. The binary distribution function $\varphi(\mu)$ 
is a function of the ratio of the secondary to the combined mass of the binary system $\mu$, which is normalised on the interval $\mu\in [0,0.5]$.

\begin{figure}
\resizebox{\hsize}{!}{
\includegraphics{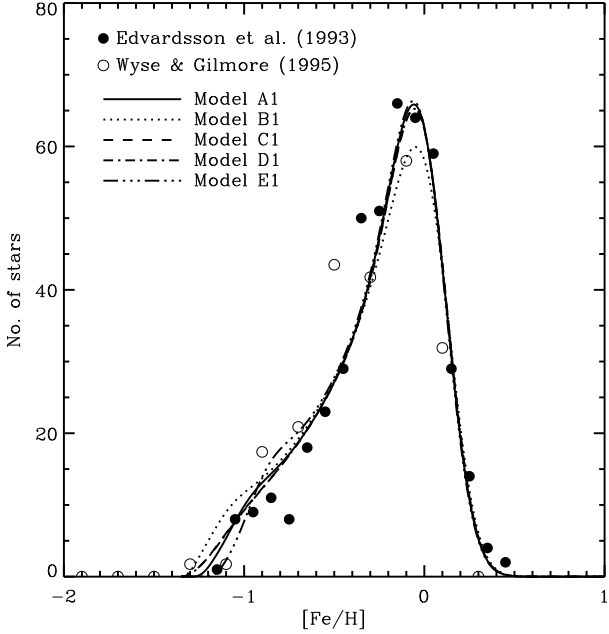}}
\caption{\label{gdwarf}
Predicted metallicity {distribution} for the solar neighbourhood {G-dwarfs} according to {models} A1-E1 convolved with a gaussian with
$\sigma = 0.15$ to simulate observational errors. }
\end{figure}

\section{Abundance data}
\label{data}
{For the Galactic HII regions, the oxygen and carbon abundances and the distances relative to the Sun,} were taken from Esteban et al.~(2002; 2005).
For the extragalactic {HII regions}, data were taken from Bergvall (1985), Garnett et al.~(1995; 1997; 1999), Kobulnicky et al.~(1997) and
Izotov \& Thuan (1999). In all cases the spectra are corrected for extinction and underlying absorption in the Balmer lines.
The carbon abundances by Esteban et al. (2005) are also corrected for the presence of a carbon dust component.

Stellar abundance data were compiled from several different studies. For the solar neighbourhood, the data were compiled from several authors
\cite{Bensby06,Ecuvillon04,Fabbian09,Gratton00,Gustafsson99,Israelian04,Jonsell05,Spite05}, while the Cepheid abundances in Fig. \ref{grad1}
were taken from the works by Andrievski et al.~(2002a; 2002b; 2002c; 2004) and Luck et al. (2003; 2006).
\nocite{Andrievski02a, Andrievski02b, Andrievski02c, Andrievski04, Luck03, Luck06}
The observed G-dwarf distributions used in this paper were taken from Edvardsson et al. (1993) and Wyse \& Gilmore (1995).

{Results} for HI and H$_2$ (CO) {were taken} from Dame (1993). \nocite{Dame93} To obtain a consistent and, possibly, more correct
set of data for the hydrogen distributions, the conversion from CO flux to H$_2$ surface density has been redone using the oxygen-dependent
calibration by Wilson (1995), {assuming} the O/H-gradient derived by Esteban et al.~(2002; 2005). \nocite{Wilson95}

\begin{figure}
\resizebox{\hsize}{!}{
\includegraphics{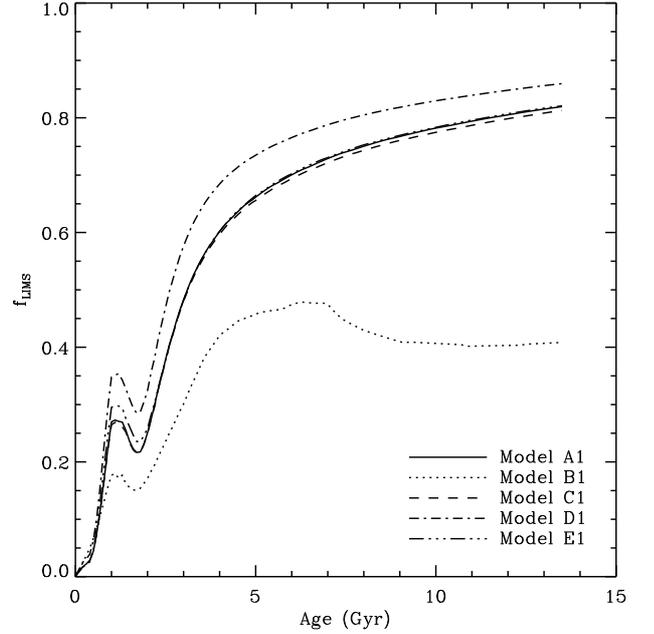}}
\caption{\label{cfrac}
Cumulative fraction of carbon due to LIM-stars as a function of time at the solar vicinity, according to {model} A1-E1 in this paper.}
\end{figure}

\begin{figure*}
\sidecaption
\includegraphics[width=12cm]{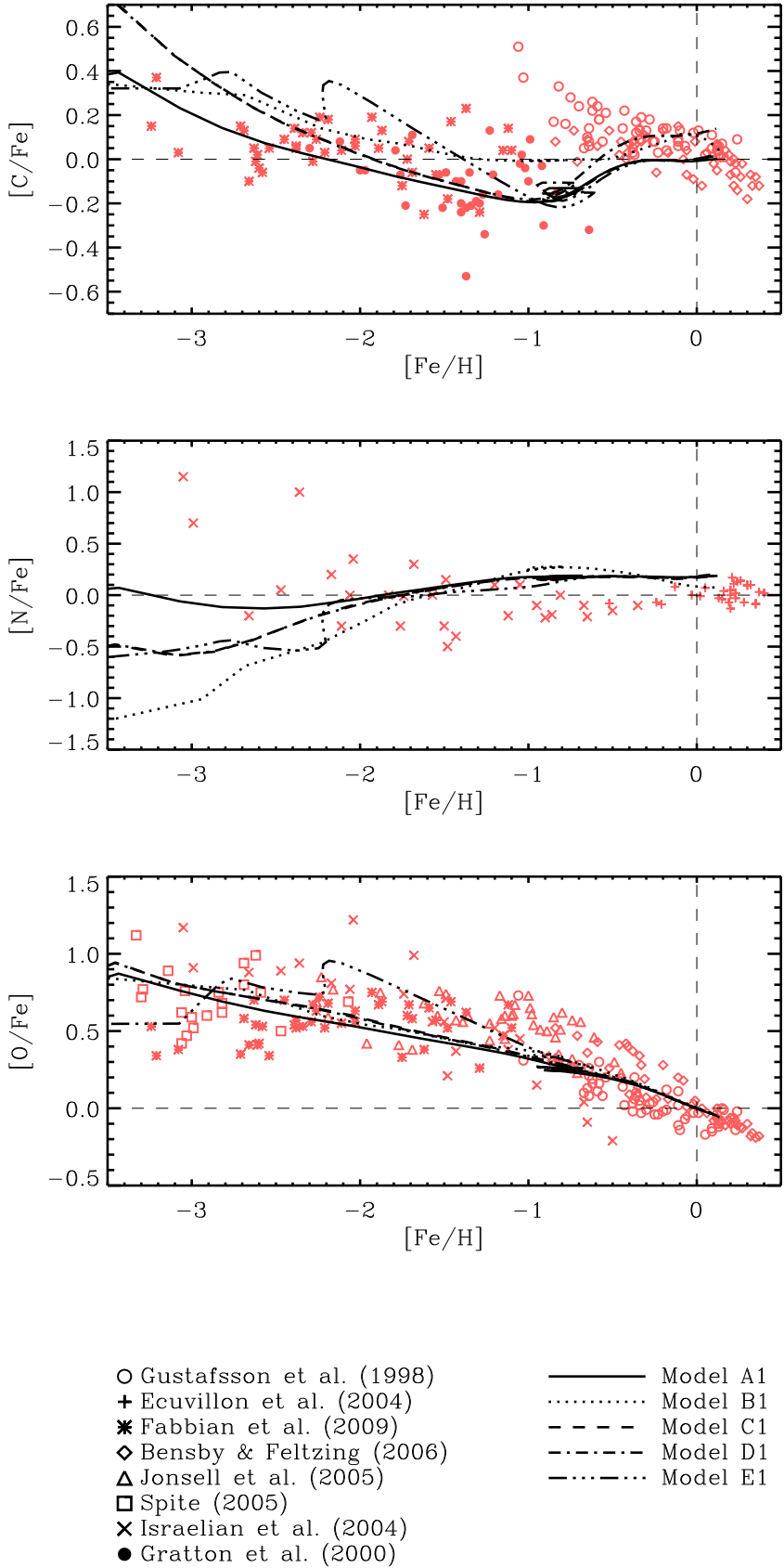}
\caption{\label{solar}
Stellar abundances of carbon, {nitrogen} and oxygen, relative to iron in the solar neighbourhood. All lines (models) and symbols (observations)
are explained in the figure. {Data from Fabbian et al. (2009a) is that derived accounting for H-collisions.}
The thin dashed lines marks the solar values. Models with the second type of star formation law (A2-D2) are not shown,
but yield very similar results. The discontinuity {in the model tracks seen in the carbon abundance plot} around [Fe/H] = -1 is due to the onset
of disc formation, i.e., the second infall episode.}
\end{figure*}

  \begin{table*}
  \begin{center}
  \caption{\label{solarcomp2} Predicted and observed quantities of the solar neighbourhood. {$\dot{\Sigma_\star}$ and $\dot{\Sigma}_{\rm inf}$
           denote star-formation rate and infall rate, respectively.}}
  \begin{tabular}{llllllllllll}
  \hline
  \hline
  Quantity                               & Unit                           & A1   & B1   & C1   & D1   & E1   & A2   & B2   & C2   & D2   & Observed\\
  \hline
  SN I rate                              & $10^{-3}$ pc$^{-2}$ Gyr$^{-1}$ & 1.11 & 1.44 & 1.15 & 1.12 & 1.09 & 1.00 & 1.54 & 1.00 & 1.00 & -\\
  SN II rate                             & $10^{-3}$ pc$^{-2}$ Gyr$^{-1}$ & 3.81 & 3.75 & 3.87 & 3.89 & 3.69 & 3.17 & 3.82 & 3.17 & 3.20 & -\\
  SN II/SN I                             & -                              & 3.43 & 2.60 & 3.37 & 3.47 & 3.39 & 3.17 & 2.48 & 3.17 & 3.20 & $\sim$ 3.1\\
  $\dot{\Sigma}_\star(R_\odot, t_0)$     & $M_\odot$ pc$^{-2}$ Gyr$^{-1}$ & 3.25 & 2.82 & 3.30 & 3.33 & 3.17 & 2.82 & 2.89 & 2.82 & 2.87 & 2 - 5\\
  $\Sigma_{\rm gas}(R_\odot, t_0)$       & $M_\odot$ pc$^{-2}$            & 7.81 & 7.98 & 7.95 & 7.93 & 7.68 & 7.44 & 8.04 & 7.44 & 7.47 & 8 $\pm$ 5\\
  $\dot{\Sigma}_{\rm inf}(R_\odot, t_0)$ & $M_\odot$ pc$^{-2}$ Gyr$^{-1}$ & 1.44 & 1.44 & 1.44 & 1.44 & 1.44 & 1.30 & 1.30 & 1.30 & 1.30 & 0.5 - 5\\
  \hline
  \end{tabular}
  \end{center}
  \end{table*}

  \begin{table*}
  \begin{center}
  \caption{\label{solarcomp} Solar abundances by mass predicted {for 4.56 years ago} by the models and compared with the current solar composition
  {derived from observations}.}
  \begin{tabular}{lllllllllllll}
  \hline
  \hline
  & A1 & B1 & C1 & D1 & E1 & A2 & B2 & C2 & D2 & Asplund et al. (2005)\\
  \hline
  H       & 0.736             & 0.729           & 0.736           & 0.736           & 0.736
          & 0.736             & 0.729           & 0.736           & 0.736           & 0.739 \\
  He      & 0.253             & 0.260           & 0.253           & 0.253           & 0.253
          & 0.253             & 0.260           & 0.253           & 0.253           & 0.249\\
  C       & $2.18\,10^{-3}$   & $2.77\,10^{-3}$ & $2.18\,10^{-3}$ & $2.85\,10^{-3}$ & $2.18\,10^{-3}$
          & $2.18\,10^{-3}$   & $2.69\,10^{-3}$ & $2.20\,10^{-3}$ & $2.86\,10^{-3}$ & $2.18\,10^{-3}$\\
  N       & $9.32\,10^{-4}$   & $7.54\,10^{-4}$ & $9.24\,10^{-4}$ & $9.43\,10^{-4}$ & $9.32\,10^{-4}$
          & $9.22\,10^{-4}$   & $7.38\,10^{-4}$ & $9.22\,10^{-4}$ & $9.47\,10^{-4}$ & $6.23\,10^{-4}$\\
  O       & $5.41\,10^{-3}$   & $5.41\,10^{-3}$ & $5.41\,10^{-3}$ & $5.41\,10^{-3}$ & $5.41\,10^{-3}$
          & $5.41\,10^{-3}$   & $5.41\,10^{-3}$ & $5.41\,10^{-3}$ & $5.41\,10^{-3}$ & $5.41\,10^{-3}$\\
  Fe      & $1.17\,10^{-3}$   & $1.17\,10^{-3}$ & $1.17\,10^{-3}$ & $1.17\,10^{-3}$ & $1.17\,10^{-3}$
          & $1.17\,10^{-3}$   & $1.17\,10^{-3}$ & $1.17\,10^{-3}$ & $1.17\,10^{-3}$ & $1.17\,10^{-3}$\\
  \hline
  \end{tabular}
  \end{center}
  \end{table*}

\section{Results and discussion}
\label{results}
Using the framework described {in Sect.~\ref{cem} in association with a {variety} of nucleosynthetic prescriptions, we computed the following
five types of CEMs for the Milky Way:}
\begin{itemize}
\item [(A)] HM-star yields by WW95 \nocite{WW95} {and LIM-star yields} by Gavil\'an et al. (2005) \nocite{Gavilan05} {and Gavil\'an et al.
            (2007), \nocite{Gavilan07} the latter being for low-metallicity.}
\item [(B)] HM-star yields by Portinari et al. (1998) and
            LIM-star yields from van den Hoek \& Groenewegen (1997).  \nocite{Hoek97}
\item [(C)] As in model A, except that for HM-stars the carbon yields at zero metallicity ($Z = 0$) are increased by 50\%.
\item [(D)] As in model C, except that the LIM-star carbon yields below $Z = 0.01$ are increased by 50\% in an attempt to reproduce the
            highest {carbon abundances derived for objects having metallicities around solar}.
\item [(E)] Yields as in model A but now with an evolving IMF according to Eq. (\ref{evolimf}).
\end{itemize}
Case A represents a case where the LIM-stars are producing most of the carbon, while case B represents a case where the carbon is to a large
extent produced in HM-stars, similar to the scenario suggested by Gustafsson et al. (1999). Case C is an attempt to explain the {C/O trend
obtained by Akerman et al. (2004) and Fabbian et al. (2009a) from observational data of stars at low metallicity.} The modified LIM-star yields in
Case D aims at explaining the super-solar carbon abundances observed in the solar neighbourhood \cite{Gustafsson99,Esteban05,Carigi05,Bensby06}.
Finally, case E is introduced as an alternative solution to the early C/O evolution, where instead of modifying the yields for the first
generation of stars, a top-heavy IMF is used at low gas densities. In all models the iron yields of WW95 at $Z = 0$ have been reduced by a factor
of five in order to obtain O/Fe-ratios consistent with those found in very metal-poor stars (see also the discussion in Sect. \ref{iron}). That
the yields of WW95 may lead to an over-production of iron, at low metallicity in particular, was pointed out already by Timmes et al. (1995).

\begin{figure*}
  \begin{center}
  {
  \includegraphics[width=18.5cm]{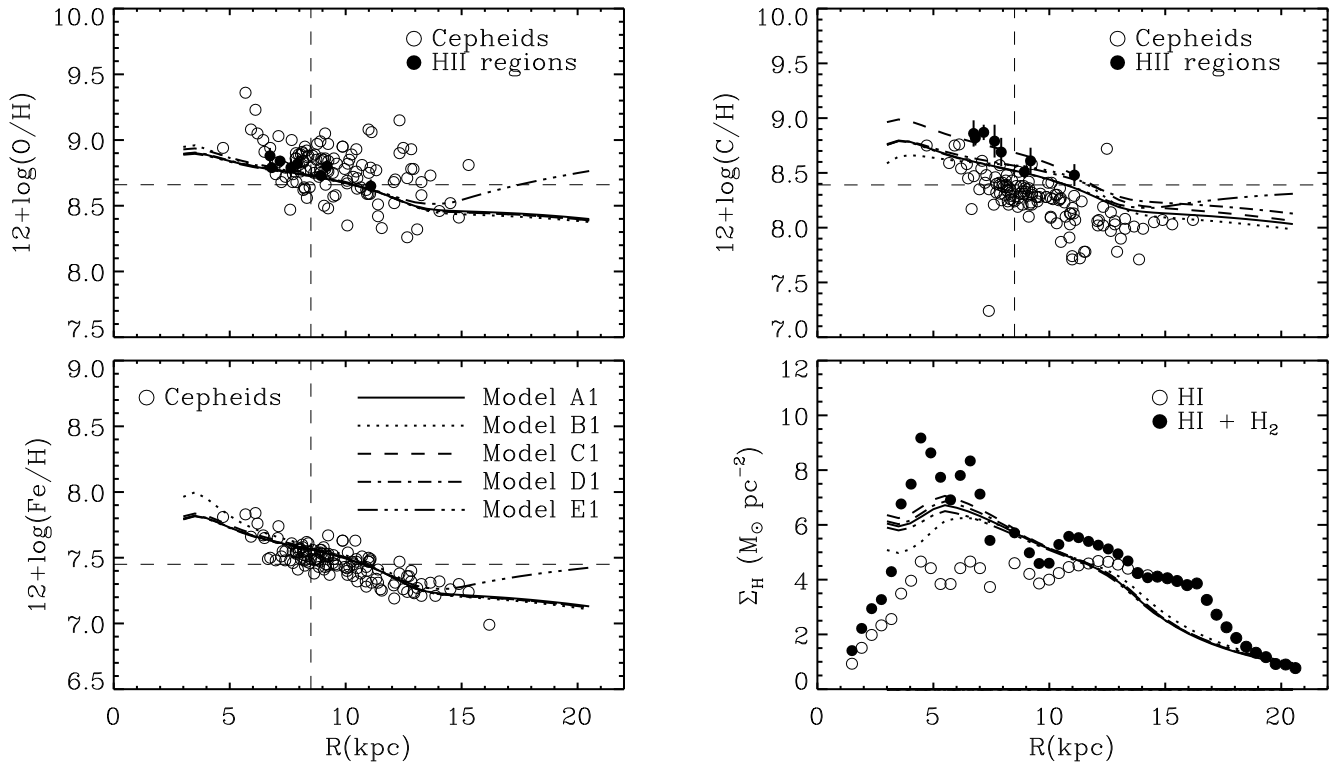}
  }
  \end{center}
  \caption{\label{grad1}
  Predicted and observed radial trends for O/H, C/H (upper panels), Fe/H and the surface density of hydrogen (lower panels) {in} the Milky Way
  for models computed using the star formation prescription of type 1. {Solid circles} show abundances in HII regions, {while empty} circles show abundances
  in Cepheids. The {vertical and horizontal} thin, dashed lines indicate the solar values according to Asplund et al. (2005).}
  \end{figure*}

The model parameters were adjusted {to achieve good agreement with the abundance data described in Sect. 3, i.e., either attempting to reproduce
the radial trends of elemental abundances} (star formation type 1, referred to as models A1-E1, see Fig. \ref{grad1}) {or of the molecular
hydrogen density as derived by \cite{Dame93} from observations of the outer parts of the Galaxy} (star formation type 2, referred to as models A2-D2,
see Fig. \ref{grad2}). Other properties, such as {the solar abundance pattern}, the observed G-dwarf distribution and {basic solar
neighbourhood quantities} were also used as constraints in both cases (see Table \ref{solarcomp2}, Table \ref{solarcomp} and Fig. \ref{gdwarf}).

\subsection{Solar neighbourhood trends}
\begin{figure*}
\sidecaption
\includegraphics{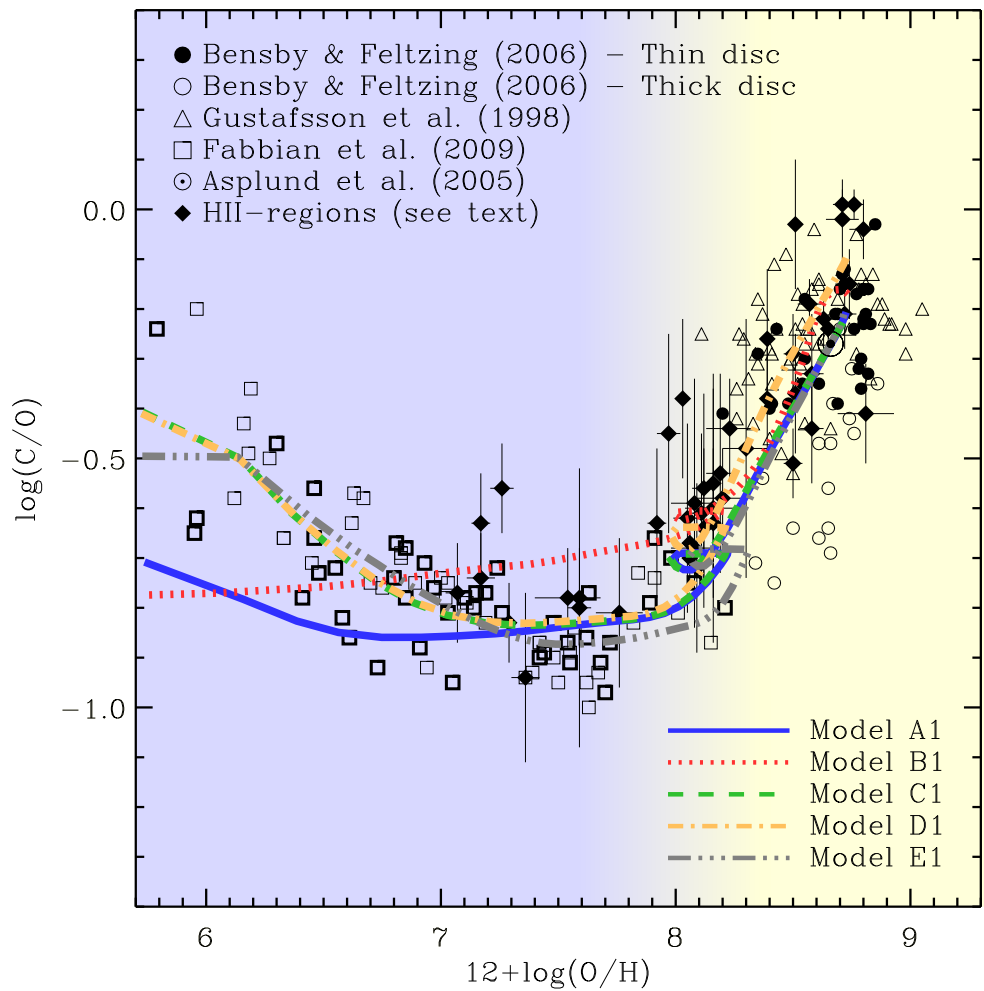}
\caption{\label{CO}
Carbon abundance relative to oxygen. Note that Galactic and extragalactic HII-regions
appear to follow a trend similar to that of the solar neighbourhood. The different lines have the same meaning as in {Fig. 4-8}, i.e.,
they correspond to {models} A1-E1.
The yellow-shaded (bright) area indicates roughly where C stars dominate the carbon enrichment of the ISM according to {models A1, C1 and E1}.
Thin squares show
data from Fabbian et al. (2009a) {assuming no hydrogen collisions in their abundance derivations}, while thick squares show the same data set,
{but accounting for hydrogen collisions. The quantitative model results are uncertain during the first few time steps, which is the reason the
plot is cut at $\log({\rm O/H}) = 5.7$ even though the data by Fabbian et al. (2009a) includes a couple of stars at even lower metallicity.
The uncertainty is a consequence of the numerical methods used and the fact that simulated abundance ratios may, in that early phase, change
quite significantly due to the step sizes (linearly interpolated) in tabulated input data.}}
\end{figure*}

For the chemical evolution of the solar neighbourhood ({defined, in the models presented here}, as the evolution at galactocentric radius
$R = 8.5$ kpc) {prescription of type 1 and 2 for star formation} provide very similar results. Hence, only models A1-E1 will be discussed here.
Model A1 provides satisfying results for the O/Fe trend and also a quite good agreement with the observed C/Fe trend, except at disc-like
metallicities ([Fe/H]$>-1$), where observations suggest a higher {C/Fe} (see Fig. \ref{solar}). However, model {A1} fails to explain the C/O trend
shown by metal-poor stars (Fig. \ref{CO}). Model B1, in which a considerably larger fraction (compared to model A1, see Fig. \ref{cfrac}) of carbon
comes from HM-stars {seems} to slightly over-produce carbon relative to iron, compared to corresponding {ratios derived from observations}
for the solar neighbourhood and, just as model A1, it cannot reproduce {the} C/O trend at low metallicity.

In models C1 and D1, the carbon production at zero metallicity is significantly increased, which may provide a solution to the
carbon-enrichment problem at low metallicities, as suggested by Akerman et al. (2004) and Carigi et al. (2005). The $Z=0$ yields for HM-stars
by Chieffi \& Limongi (2004), \nocite{Chieffi04} provide a C/O-ratio that is consistent with that {derived from} observations at very low
metallicity, but it is roughly the same also at moderately low metallicities where {observations instead suggest that} C/O {becomes} much lower.
If the carbon yields {were} indeed higher than according to WW95, {but} only at $Z = 0$, it {would pose} an interesting problem for
nucleosynthesis modelling: why {do massive}, essentially metal-free, stars produce much more carbon than more metal-rich ones {of similarly
high mass?} 

In the present paper, effects of inhomogeneities and contributions from pair-instability supernovae during the earliest evolution of the Galaxy are 
not studied, although the latter may still play an important role also for the C/O-ratio \cite[cf.][]{Karlsson05,Karlsson08}. The total carbon yield
may thus be significantly higher at early times, as in models C1 and D1. Despite the high carbon yields at $Z=0$, also models C1 and D1 are reasonably
consistent with the observed C/Fe vs. Fe/H-trend at low metallicity (again, see Fig. \ref{solar}). Model D1, however, is the only model that
seems to explain the highest C/Fe-ratios found at the high metallicity end.

The solar abundance pattern (taken to be the composition of the ISM at $R=R_\odot$, 4.56 Gyr ago) is well reproduced by all models -- except {for
an over-production of nitrogen in general and for models} D1 and D2, where the carbon abundance is about 30\% too high (see Table \ref{solarcomp}) -- 
but it should be noted that all model parameters are calibrated such that solar abundances and the peak of the G-dwarf distribution (Fig. \ref{gdwarf})
are reproduced. {For disc-like metallicities, the apparent over-production of nitrogen by the models seen in Fig. \ref{solar} may be a result of
the LIM-star yields chosen for this study in combination with the prescription for star formation. Both van den Hoek \& Groenewegen (1997) and
Gavil\'an et al. (2005) predict high nitrogen yields. In the low-metallicity end, however, the models seem to underproduce nitrogen, although {the}
high nitrogen abundances derived from observations of {some} halo stars are probably the exception rather than the rule \cite{Israelian04}.}

Model E1, in which an evolving IMF was used, provides an alternative solution to the early carbon-enrichment problem. Using the yields by WW95,
without modifications, it reproduces the C/O vs. O/H trend and simultaneously the C/Fe vs. Fe/H at low metallicity fairly well.
The reason why an evolving IMF can solve C/O vs. O/H problem using the yields by WW95, is that the C/O-ratio of the
ejecta is much higher for stellar masses around $M_\star = 10 - 15 M_\odot$ than for more massive stars. Hence, if the IMF is peaking at
intermediate stellar masses rather than at subsolar masses, the IMF weighted C/O-ratio is roughly 0.5 dex higher than it would be otherwise.
Whether an evolving IMF {is} more likely than a scenario with high carbon yields at $Z=0$ is matter of {debate}, but several recent
studies suggest that in the early Universe, star formation took place according to an IMF that was more top-heavy than at present time
\cite[see, e.g.,][and references therein]{Tumlinson06, Dave08, vanDokkum08}.

\subsection{Abundance gradients}
The Fe/H-gradients predicted by the models all agree quite well with the observed
Fe/H-gradient obtained from Cepheids (see Fig. \ref{grad1}). A similar radial trend was also found by Nordstr\"om et al. (2004)
\nocite{Nordstrom04} for young stars at galactocentric distances between 6 and 10 kpc.

\begin{figure*}
  \begin{center}
  {
  \includegraphics[width=18.5cm]{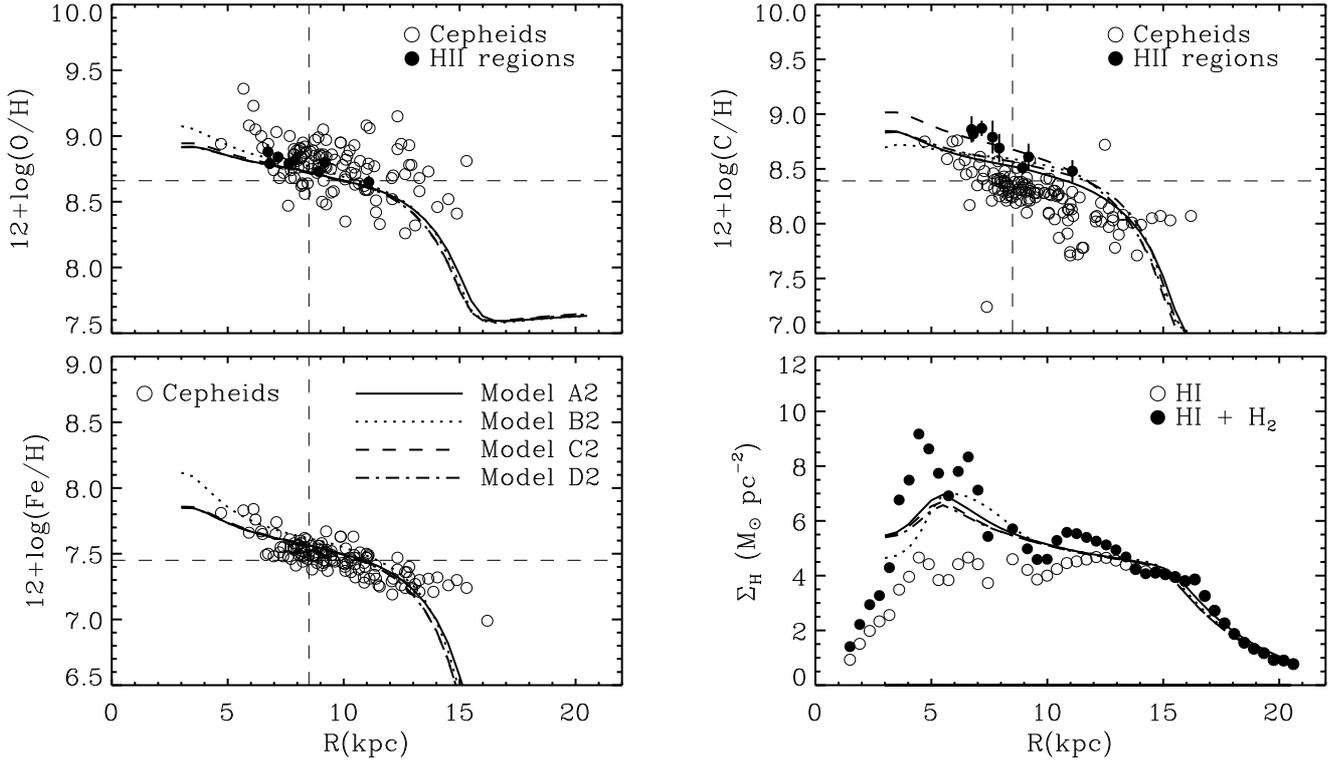}
  }
  \end{center}
  \caption{\label{grad2}
  Same as Fig. \ref{grad1}, but for models computed using the star formation prescription of type 2.}
  \end{figure*}

The results regarding the C/H-gradient are inconclusive. The ISM abundances \cite{Esteban05,Carigi05} are quite high and in reasonable
agreement with all models, although {models C1 and C2} fit these data better {(see Fig. \ref{grad1} and \ref{grad2})}. On the other hand, 
the Cepheid carbon abundances are significantly lower, {which better matches, at solar galactocentric distance, the current} solar abundance 
of carbon. {However, the} ranges of abundances seen in 
in the solar neighbourhood are rather wide and the Sun may therefore not represent the typical composition of the ISM in the solar neighbourhood at 
the time of formation of the Sun. This suggests that {calibrating the parameters of the CEMs against the current solar abundance pattern may cause 
the present-day C/H predicted by these models to end up} above (or below) that measured in young stars, {which one may assume reflects the C/H of the ISM}. 
It might be that the solar abundance pattern should not at all be used as a constraint on CEMs {for the epoch when the Sun was formed}, since its 
chemical composition (relative to iron) may depart from those of most solar-type stars of similar ages and orbits 
\cite[for further discussion, see][and references therein]{Gustafsson08}, {in fact even from those of solar twins 
\cite{Melendez09a,Melendez09b,Nordlund09}. 
In any case, the abundance} difference between Cepheids and HII-regions {still lacks} a proper explanation.

Models A2-D2 were designed to fit the hydrogen distribution \cite[according to the results of][]{Dame93} as well as possible. The good agreement is
mainly a result of the assumption that there {exists} a true star-formation threshold in the disc, below which star formation will essentially
cease completely \cite[in principle identical to the assumption made by, e.g.,][]{Chiappini97,Chiappini01}. However, the fit to the hydrogen
distribution is obtained at the expense of a good fit to the abundance gradients {derived from observations} in the outer parts of the Galaxy.
As shown in Fig. \ref{grad2}, the threshold causes the abundance gradients {predicted by the models} to fall-off steeply at galactocentric
distances beyond 13 kpc, which is the part of the disc where the gas density in the model never reaches the threshold value. {On the basis of
this disagreement with C, O and Fe abundance gradients derived from observations}, it is therefore not likely that star formation will actually
cease completely when the gas density drops below the critical density.

The somewhat poorer agreement between models A1-E1 and the observed hydrogen distribution should not be seen as a major problem. In the
present study, the possibility of having a slow radial gas flow has not been studied. Such a flow has not been observed, but even a very slow
radial gas flow can alter the hydrogen distribution quite significantly, provided enough time {is available} for the process. Hence, the
hypothesis presented here {via models A1-E1 (i.e.,} that there are two modes of star formation rather than an actual threshold) is not excluded
by the slight inconsistency {with} the HI observations of the outer disc.

\nocite{Garnett95,Garnett97,Kobulnicky97,Izotov99,Esteban02,Esteban05}

\subsection{The inverse chemical evolution problem}
The model of galaxy evolution described in Section \ref{cem} has many essentially free parameters. Nonetheless, models of this type have proven
successful in several ways, e.g., by being able to simultaneously reproduce observational constraints such as the G-dwarf distribution, basic
abundance patterns and the age-metallicity relation for stars in the solar neighbourhood, although the latter must be regarded as a rather
uncertain constraint \cite[see][]{Feltzing01}. However, due to the many parameters, it is not obvious that one may take {abundances derived from 
observations} and, 
via CEMs, derive what stellar sources (types of stars) contribute the carbon and how the production of carbon in these stars may vary with 
metallicity. In principle, one has a really non-trivial {\it inverse problem} to solve, i.e., finding the stellar yields from observed abundance 
patterns. Basic requirements for such an attempt are that a radically different star formation history should not alter the evolution of 
abundance ratios too much and that the IMF is universal and non-evolving {or that it is possible to specify how the IMF is evolving}. 

The modified sets of yields (C and D) were designed to improve the agreement between models and observations at different metallicities. 
However, for stellar abundances in the solar neighbourhood, low metallicity often means that the star belongs to the halo/thick disc population. 
The halo/thick disc population has, of course, a different formation history and therefore the evolution of elemental ratios may be slightly 
different. Several, very different, infall-time-scales and star formation prescriptions were tested, but such variations of the CEM {did} not 
change the predicted solar neighbourhood abundance {pattern much}. The two different prescriptions for star formation used in this paper
(type 1 and 2) provide very similar abundance trends for the solar neighbourhood {with only minimal changes to other model parameters}. In 
fact, the IMF and the stellar yields, or the IMF-weighted {\it integrated} yields, turn out to be the most critical ingredients. Changes in the 
integrated yields will dominate over reasonable variations of all other parameters. It is therefore possible to use CEMs to analyse how 
stellar yields may vary with stellar mass and metallicity to be consistent with observations, provided that the IMF {(and its possible 
evolution) is known.} 

It seems that the carbon abundances in stars at low metallicity in the solar neighbourhood cannot be explained by a standard Galactic evolution 
scenario and commonly used stellar yields (cf. model A1 and B1). Consequently, modifying the yields to obtain better agreement with observations, 
as a solution to the inverse chemical evolution problem, may therefore be a justified measure. However, as model E1 shows, abandoning the idea of
a universal {and non-evolving} IMF, may just as well be the solution to the carbon-enrichment problem for 
the early stages of Galactic evolution. Observed abundances (except those of extremely metal-poor stars, perhaps) reflect the enrichment from a
stellar population, and therefore one must be careful suggesting that abundances derived from observations pose a challenge to
nucleosynthesis models and stellar evolution theory. We are left with two main options: (a) the carbon yields of essentially
metal-free HM-stars must be much higher than predicted by WW95, or (b) the IMF evolves with time and/or is not completely universal. A
necessary prerequisite for actually solving the carbon enrichment problem is thus to establish which hypothesis holds. In case both (a)
and (b) is true, this would require a much less top-heavy IMF at early times, which may not be consistent with observational constraints of some 
extremely metal-poor halo stars.

The carbon enrichment at later stages (thin disc evolution) can also be explained in two ways, either: (a) the carbon contribution from HM-stars
is {metallicity-dependent} in such a way that it increases significantly with metallicity during the thin disc evolution, or (b) the rising trend
is due to a significant contribution from the long-lived stars (with masses $\le 2M_\odot$) that becomes important only after a few Gyr. Scenario 
(a) is the one advocated by Gustafsson et al. (1999), while (b) represents the picture favoured by several authors 
\cite[e.g.,][]{Chiappini03,Akerman04,Gavilan05}, in 
which up to 80\% of the carbon in the present-day ISM may be due to LIM-stars (see Fig. \ref{cfrac}). It is virtually impossible to distinguish 
between (a) and (b) just by considering abundance data, but the fact that carbon and iron seem to be {\it released} to the ISM on similar 
time-scales, {suggests} that these two elements originate from stars that {\it evolve} on similar time-scales. If LIM-stars produce most of the 
carbon and supernovae Type Ia produce the bulk of iron in the Galaxy, the rather flat trend of C/Fe with Fe/H would be explained naturally
\cite{Bensby06}. If, {instead,} scenario (a) is correct, it would require that the metallicity dependence of the carbon yields of HM-stars is 
fine-tuned {with the SNIa-produced Fe} in such a way that {both the flat C/Fe vs. Fe/H trend and the rising C/O vs. O/H trend are reproduced.
Thus, while (a) is a plausible hypothesis, scenario (b) is simpler in that it} does not require that the yields have any specific metallicity 
dependence; it only requires that LIM-stars produce most of the carbon, so that the carbon enrichment of the ISM is delayed. 

An interesting fact, which {deserves} further study ({beyond the scope of this paper}), is evident from data plotted in Fig. \ref{CO}. 
Extragalactic HII-regions show C/O-ratios which line up along the same trend with O/H as the stars in the solar neighbourhood: metal-poor (low-mass)
galaxies show low carbon abundances, while more metal-rich galaxies show considerably higher carbon abundances, relative to oxygen. This may, in 
principle, be consistent with both scenario (a) and (b) {just discussed for the thin disc. Carbon production in the Galactic disc appears to be
coupled with the age/metallicity of the stellar population. Similarly, it seems that the carbon abundance in extragalactic HII-regions is coupled 
with the age of the dominant stellar population or the overall metallicity.} However, if (b) is the correct scenario, metal-poor galaxies must be 
dominated by young stellar populations, so that the majority of LIM-stars have not yet evolved off the main sequence, {thus locking carbon,
without significantly enriching the ISM. Nevertheless, the light distribution of HII/BCG galaxies reveals the presence of a significant low
surface-brightness population} of old stars \cite{Telles97,Kunth00}, although it is difficult to determine the mass fraction of {such an underlying 
population.} If scenario (b) can be confirmed, metal-poor galaxies {are very likely to} have formed most of their stars quite recently, 
{since a significant old population would raise the C/O-ratio due to enrichment from LIM-stars.} 

It is often suggested {\cite[see][and references therein]{Kunth00}} that low-mass, metal-poor galaxies are deficient in metals due to significant 
galactic winds and may therefore have undergone several star-forming episodes in the past. Galactic winds may be responsible for maintaining low 
over-all metallicities, but cannot easily explain low C/O-ratios {unless the winds are extremely selective (oxygen remains while carbon is 
expelled from the galaxy) in scenario (b)}. {The fundamental question is then whether LIM-stars are (or not) the main contributors to carbon
nucleosynthesis in metal-poor galaxies, and if galactic winds may maintain low over-all metallicities even after a sequence of star-forming
episodes.}

\subsection{Uncertainties in iron production}
\label{iron}
Several uncertainties have not been addressed in detail here. One, particularly important, is the production of iron, which cannot 
be well-constrained for two reasons: the location of the mass cut (that divides the part of the star that collapses in the remnant from that 
which is expelled) in supernova nucleosynthesis models is a free (although restricted) parameter, and the origin of supernovae Type Ia (SNIa) 
is not fully understood. 

The iron yields by WW95 show a dependence on metallicity (possibly an effect of how the mass 
cut was chosen), which may not be realistic. At $Z=0$ the IMF-weighted WW95 yields show {an} O/Fe-ratio that is much lower than the {ratios 
derived from observations} in very metal-poor stars ({[O/Fe]$\sim 0.8$ when} [Fe/H]$<-3$, see Fig. \ref{solar}), which is the reason why the iron 
yields at $Z = 0$ have been lowered to 1/5 of their original values in the models presented in this paper. {However, non-local thermal equilibrium
(non-LTE) effects on oxygen lines may explain the O/Fe discrepancy. Including corrections for non-LTE effects, smaller
[O/Fe] values are derived from observations, so the [O/Fe] may actually agree quite well with WW95 \cite{Fabbian06, Fabbian09b}.}

The fact that the iron yields of HM-stars are rather uncertain, and models of the evolution of SNIa rates cannot be very well constrained from 
observations, makes the common practice {of using} iron as the reference element (representing the over-all metallicity) somewhat questionable 
when comparing data and models. Hence, it makes more sense to calibrate CEMs primarily against abundances relative to, e.g., oxygen or other common 
elements for which the enrichment essentially follows the star formation rate. Therefore, focus is on reproducing C/H, O/H and C/O, rather than C/Fe 
and O/Fe, in the present study.

\subsection{Uncertainties in carbon production: effects of carbon star mass loss}
There are many assumptions and physical prescriptions that may affect the nucleosynthesis in stellar evolution models. For LIM-stars the duration of 
the AGB phase and the number of thermal pulses is almost uniquely determined by the mass-loss rate. The evolution of the internal structure depends 
on the mass-loss rate as well \cite{Blocker95}, which in turn affects the fundamental stellar parameters. Since the mass-loss rate depends on these 
stellar parameters, there will be a feedback, which means that the mass-loss prescription put into a stellar evolution model is critical, and it is 
indeed well-known that changing the mass-loss prescription can have profound effects on the yields of AGB stars (see, e.g., van den Hoek
\& Groenewegen 1997). More precisely, the sum of carbon and nitrogen synthesised in LIM stars is largely controlled by the mass-loss rate on the 
AGB. 

In the work of Gavil\'an et al. (2005) it was assumed that stars of low metallicity have smaller total radii (yielding higher surface gravity) 
and thus less effective mass-loss, which {increases} the lifetime of the AGB phase and {allows} these stars to experience several more
dredge-up events where carbon is mixed into the envelope. {At late stages, as metallicity increases,} the carbon production {declines}, and 
secondary nitrogen production becomes increasingly significant \cite[see also][for further details]{Buell97}. But is it correct to assume that  
low metallicity means less effective mass-loss {relative to that of solar metallicity stars}?

Mattsson et al. (2008) \nocite{Mattsson08} have shown that, according to theoretical models, the overall metallicity is not affecting dust-driven 
mass-loss from C-stars much at all. It is mainly the abundance of carbon, available for dust formation, that has a significant effect.
The dust-driven mass-loss takes place mainly during the late stages of evolution, where thermal pulses and dredge-up events dominate the
evolution. There is no particular reason to assume that low metallicity C-stars will experience more thermal pulses and dredge-up events before the
termination of the AGB. Furthermore, Mattsson et al. (2010a) present a state-of-the-art model of the late stages of evolution of a $2M_\odot$-star
using a new detailed mass-loss prescription based on the results by Mattsson et al. (2010b) and a new set of low-temperature opacity coefficients,
in which effects of the abundances of carbon and oxygen were taken into account. This resulted in a significantly lower effective temperature and
the development of a very pronounced so-called superwind, shortly after the star had become carbon rich, which terminated the C-star evolution after 
only few thermal pulses. The amount of carbon that can be dredged up and expelled by the stellar wind is therefore limited. No final word yet, but 
these recent findings seem to restrict the carbon production in LIM-stars {by means of a self-regulating process, where the mass-loss rate 
increases with every dredge-up event associated with each thermal pulse.}

\section{Conclusions and final remarks}
CEMs for the Milky Way have been presented. Four different nucleosynthesis prescriptions were used, where two contain {\it ad hoc} modifications to 
meet the {abundance trends of C/O vs. O/H and C/Fe vs. Fe/H derived from observations}. According to these CEMs one may conclude the following:
\begin{itemize}
\item Carbon is being released to the ISM on {time-scales} comparable to that of iron, {which} is produced {mainly} by
      supernovae Type Ia.
\item An evolving IMF, being top-heavy (favouring the formation of HM-stars) during the early stages of Galactic evolution, can explain the C/O vs.
      O/H trend seen in the {abundances derived} by Fabbian et al. (2009a), without violating any other observational constraints.
\item The {C/H-gradient (or the metallicity gradient in general) derived from observations} in the Milky Way disc suggest there is a problem 
      with simple CEMs and the hydrogen distribution. Introducing a star formation threshold at $\Sigma_{\rm c} = 7 M\odot$ pc$^{-2}$ can 
      explain the {shape of the hydrogen distribution}, but such a model predicts a steep fall-off in all abundance gradients beyond a certain 
      galactocentric distance and cannot explain the flattening of the C/H and Fe/H gradients of the outer disc traced by Cepheid observations. It 
      is possible that radial gas flows in the outer disc in combination with a "low-efficiency mode" of star formation {is} responsible for the 
      abundance gradients and the hydrogen distribution. 
\item HM-star {yields of carbon and oxygen} probably have a metallicity dependence, as argued by {Gustafsson et al. (1999)}, but the 
      observed increase with metallicity of the C/O-ratio seen in disc stars could just as well be due to a delayed release of carbon from 
      LIM-stars. If the best-fit models are correct, the major source of carbon in the present-day Galaxy is the LIM-stars, providing as much as 
      80\% of the carbon to the ISM. Although HM-stars as major carbon producers cannot be excluded, the LIM-star scenario provides a simpler 
      explanation, which does not require that carbon yields depend strongly on metallicity. 
\end{itemize}
The origin of carbon remains an open question. The scenario advocated here, is that LIM-stars have relatively high carbon yields and that the winds 
from these stars dominate the carbon enrichment of the ISM. However, recent results by Mattsson et al. (2010a) suggest that since the mass-loss of a
carbon star has a steep dependence on the amount of carbon that is dredged up to the surface of the star, {the mass-loss rate
increases with with every dredge-up event, which eventually leads to the termination of the AGB phase. Hence,} the number of thermal pulses will be
very limited {due to this self-regulating  mechanism and} the carbon yield can hardly {become} as high as in models D1 and D2. 

That LIM-stars may contribute most of the carbon in the solar neighbourhood has previously been concluded by several other authors
\cite[e.g.,][]{Chiappini03,Akerman04,Gavilan05}. The C/Fe vs. Fe/H trend seen in unevolved, metal-poor stars in the solar neighbourhood suggests 
that the contribution from HM-stars at low metallicity must be {limited, unless these stars also have an enhanced iron production}, 
and {the} contribution {has to increase} dramatically with metallicity to explain the sharp up-turn in C/O vs. O/H, seen in Fig. \ref{CO}. 
If HM-stars {were} the main contributors, then the carbon yields of these stars (around and above solar metallicity) must be even larger than 
the oxygen yields. Maeder (1992) \nocite{Maeder92} and Portinari et al. (1998) have computed models where such results were obtained, and more 
recent work {seems} to give similar results (see, e.g., Meynet \& Maeder 2002; 2003; 2005). But models of HM-star evolution are also riddled 
by several uncertainties.  

Finally, returning to the connection between the cosmic carbon abundance and carbon-based life, one may notice that in the outer parts of the 
Milky Way disc, where the peak of star formation was reached quite recently (or has not yet been reached), there has not been enough time for the 
bulk of LIM-stars (especially those with main-sequence masses below $\sim 2M_\odot$) to evolve into mass-losing C-stars, {or} the metallicity 
is not large enough {if the HM-star carbon production scenario is instead the correct one.}
Are the outer parts of the Galaxy therefore excluded as a possible environment for complex life to emerge, since there has been no 
significant carbon enrichment until quite recently?

\begin{acknowledgements}
The author wishes to thank the referee for all valuable comments and suggestions that helped improve the clarity and readability of the paper.
Bengt Gustafsson, Nils Bergvall and Kjell Olofsson are thanked for their valuable comments on a draft version of this paper. Mercedes Moll\'a and
Marta Gavil\'an are thanked for clarifying some issues regarding the LIM-star yields.
\end{acknowledgements}

\bibliographystyle{aa}

\end{document}